# Role of spin mixing conductance in spin pumping: enhancement of spin pumping efficiency in Ta/Cu/Py structures

Praveen Deorani and Hyunsoo Yang[a]

*Department of Electrical and Computer Engineering, National University of Singapore, 117576, Singapore*

From spin pumping measurements in Ta/Py devices for different thicknesses of Ta, we determine the spin Hall angle to be 0.021 - 0.033 and spin diffusion length to be 8 nm in Ta. We have also studied the effect of changing the properties of non-magnet/ferromagnet interface by adding a Cu interlayer. The experimental results show that the effective spin mixing conductance increases in the presence of Cu interlayer for Ta/Cu/Py devices, whereas it decreases in Pt/Cu/Py devices. Our findings allow the tunability of the spin pumping efficiency by adding a thin interlayer at the non-magnet/ferromagnet interface.

[a] e-mail address: eleyang@nus.edu.sg

The prospect of using spin currents in information processing devices is a subject of great interest because of potential benefits in power dissipation and time constants[1-4]. One of the basic requirements of such applications is a large density spin current in non-magnetic materials. Spin pumping is one promising process, in which large spin current densities can be induced in a non-magnetic material (NM) attached to a ferromagnet (FM) with a precessing magnetization. Spin pumping can generate spin currents both in metals and in semiconductors, circumventing the problem of impedance mismatch. Thus, a further improvement in the efficiency of spin pumping process is important for future spintronic devices.

In recent years, the spin pumping induced spin current has been most commonly measured as an electric signal via the inverse spin Hall effect (ISHE) in NM[5-8], or as an enhancement in the Gilbert damping ($\alpha$) of the FM[9-11]. The spin pumping efficiency can be enhanced by optimizing the magnetization angle out-of-plane of the film[12], by spin wave mode selection[13], by chemical modification of FM surface by *in situ* etching, or by using $H_2SO_4/H_2O_2$ mixture for cleaning FM surface[14,15]. The contribution of surface spin waves to spin pumping signal has been discussed, and it has been shown that spin waves dramatically enhance the spin pumping induced ISHE signals[16]. The properties of the NM/FM interface, quantified by spin mixing conductance ($g_{\uparrow\downarrow}$) is also an important determinant of the spin pumping efficiency[17]. Previous reports have shown that the spin pumping efficiency can be suppressed by the presence of a thin layer of MgO[18], nano oxide[19] or titanium[20] at the NM/FM interface. The aim of this study is to observe the effect of changing the NM/FM interface on $g_{\uparrow\downarrow}$ by insertion of a Cu interlayer. Experiments have been conducted to measure ISHE induced by spin pumping in NM, as well as the enhancement of α in NM/Cu/Py devices (NM = Ta, Pt). We find that the effective $g_{\uparrow\downarrow}$ increases in the presence of Cu interlayer in Ta/Cu/Py devices, but decreases in Pt/Cu/Py devices. In addition, we report the spin diffusion length in Ta to be 8 nm and spin Hall angle to be 0.021 – 0.033.

Figure 1 (a) shows a schematic of the device with the measurement set up. The cross section of the device is shown in Fig. 1(b). The devices were fabricated using photolithography and lift-off process. First, 800 μm × 600 μm NM/Cu/$Ni_{81}Fe_{19}$ trilayers are fabricated by sputter deposition with the Cu thickness ranging from 0 to 10 nm for different devices. In all devices, the thickness of NM (Ta or Pt) layer is 10 nm and the thickness of NiFe (Py) layer is 20 nm. In the next step, a 640 μm × 630 μm × 30 nm $SiO_2$ layer is deposited to insulate the Py layer. Then, in the last step, Ta (5 nm)/Cu (150 nm) asymmetric coplanar strips (ACPS) and dc probe pads are patterned, and sputter deposited simultaneously. In the ACPS strip, the width of signal line is 60 μm, the width of ground line is 180 μm, and the signal-ground spacing is 30 μm. A bias field $H_b$ is applied along the z-direction shown in Fig. 1(b). For spin pumping induced ISHE measurements, a microwave signal of a fixed frequency is applied to the ACPS waveguide using a signal generator (SG) and a dc voltage is measured across the NM/Cu/Py trilayers as a function of $H_b$. For the evaluation of enhancement in α, ferromagnetic resonance (FMR) measurements are carried out using a vector network analyzer. All the measurements are carried out at room temperature.

Figure 2(a) shows the measured values of the electromotive force signal in Ta/Py device without a Cu interlayer. During the measurement, the signal generator applies a 15 dBm microwave at 4 GHz ($=\omega/2\pi$) to the ACPS waveguide, and the $H_b$ along the z-direction is swept across the resonance field $H_0$, such that $\omega = \gamma\mu_0\sqrt{H_0(H_0+M)}$, where $\gamma$ is the gyromagnetic ratio of a free electron, $\mu_0$ is the permeability of free space, and $M$ is the saturation magnetization of Py. The measured signal consists of ISHE in the Ta layer, and the anisotropic magnetoresistance (AMR) and the anomalous Hall effect (AHE) of the Py layer[7]. The effect of AMR or AHE can be eliminated by noting that the ISHE signal has a symmetric Lorentzian shape, whereas the signal due to AMR or AHE has an asymmetric Lorentzian shape[7,21]. Thus, the measured data are fitted by a sum of symmetric and asymmetric Lorentzian functions, $V = V_{sym}\dfrac{\Gamma^2}{\Gamma^2+(H-H_0)^2} + V_{asym}\dfrac{\Gamma(H-H_0)}{\Gamma^2+(H-H_0)^2}$ and the value of $V_{sym}$ is taken to be the spin pumping induced ISHE signal ($V_{ISHE}$). Similar measurements are carried out at 3, 4, 5, and 6 GHz, for all the Ta/Cu/Py devices, with Cu thicknesses ranging from 0 to 10 nm, and values of $V_{ISHE}$ are obtained from fitting. The resonance fields $H_0$ with the used frequency in measurements are plotted in Fig. 2(b), and is fitted with Kittel formula $\omega = \gamma\mu_0\sqrt{H_0(H_0+M)}$ to obtain $M = 1.05$ T.

The ISHE signal level is given by $V_{ISHE} \propto J_s\,\theta_{sh}\,R$,[7] where $J_s$ is the density of spin current induced from spin pumping, $\theta_{sh}$ is the spin Hall angle of NM, and $R$ is the resistance of NM/Cu/FM trilayer. The value of $R$ as measured in different devices is plotted with the Cu interlayer thickness in Fig. 2(c). The correct quantification of spin current density is given by the ratio between $V_{ISHE}$ and $R$. This ratio, as obtained from measurements in different devices and for different frequencies, is normalized with respect to that in Ta/Py device and plotted in Fig. 2(d). It is clear that the $J_s$ increases by about 2.2 times in the presence of Cu interlayer in the Ta/Cu/Py structure. It must be noted that Cu has very small spin orbit coupling and a large spin diffusion length (~ 300 nm)[22] at room temperature, therefore the effect of a thin layer (10 nm) of Cu does not have any effect in the spin Hall angle. However, the Cu interlayer changes the spin current density by changing effective interface properties, not by spin scattering.

In order to better understand the increase in $J_s$ in the presence of Cu interlayer, FMR measurements are conducted in all Ta/Cu/Py devices. The linewidth of FMR absorption spectrum is a measure of effective $\alpha$, and has been reported to be enhanced by the spin pumping effect[9,10]. The $\alpha$ can be calculated from the FMR linewidth as $\alpha = \Delta\omega/\gamma$, where $\Delta\omega$ is the full width at half maximum of the FMR spectrum and $\gamma$ is the gyromagnetic ratio of the free electron. In Fig. 2(e) the FMR linewidth measured at $H_b = 126$ Oe is seen to be largest for Ta/Cu/Py, followed by Ta/Py and Py. This result shows that the spin current density induced by spin pumping is enhanced in Ta/Cu/Py devices as compared to Ta/Py devices. The effective spin mixing conductance is related to the enhancement in effective $\alpha$ via[7,23]

$$g_{\uparrow\downarrow} = 4\pi M \gamma d_{Py} (\Delta\alpha)/(g \mu_0 \mu_B) \tag{1}$$

where $M$ is the saturation magnetization of Py layer, $d_{Py}$ is the thickness of Py layer (20 nm), $g$ is the electron g-factor (2), $\mu_0$ is the permeability of free space, and $\mu_B$ is the Bohr magneton. Figure 2(f) shows the calculated $g_{\uparrow\downarrow}$ from Eq. (1) for different Ta/Cu/Py devices. $g_{\uparrow\downarrow}$ is seen to increase up to ~ 2.1 times in the presence of Cu interlayer. The enhancement in $g_{\uparrow\downarrow}$ in Fig. 2(f) follows the same trend as that observed in spin pumping induced ISHE signals in Fig. 2(d). These results show that the Cu interlayer enhances the $g_{\uparrow\downarrow}$ at the Ta/Cu/Py interface, which results in the enhancement of the spin current density induced by spin pumping, which is given by[23]

$$J_s = -\frac{\hbar g_{r\uparrow\downarrow}\gamma^2 h_{rf}^2 \left(M\gamma + \sqrt{M^2\gamma^2 + 4\omega^2}\right)}{8\pi\alpha^2 \left(M^2\gamma^2 + 4\omega^2\right)} \frac{\lambda_{Ta}}{d_{Ta}} \tanh\left(\frac{d_{Ta}}{2\lambda_{Ta}}\right) \quad (2)$$

where $h_{rf}$ is the rf field amplitude, $\omega$ is the ferromagnetic resonance frequency ($2\pi f$), $\lambda_{Ta}$ is the spin diffusion length in Ta, and $d_{Ta}$ is the thickness of Ta layer (10 nm). Thus the result from FMR measurements matches very well with the results from spin pumping induced ISHE measurements in Ta/Cu/Py devices with different Cu thicknesses.

We further analyze our results using a simple model in which the spin mixing conductance can be written as $g_{\uparrow\downarrow} \approx \kappa k_F^2 A / 4\pi^2$,[17] where $\kappa$ is the number of conducting channels per surface atom and has a value of the order of unity, $k_F$ is Fermi wave number of NM, and $A$ is the area of interface. The values of $k_F$ for Cu and Ta can be obtained from $k_F = m_e^* V_F / \hbar = \sqrt{2 m_e^* E_F / \hbar^2}$, where $V_F$ is the Fermi velocity, $E_F$ is the Fermi energy, and $m_e^*$ is the electron effective mass. The values used for calculating $k_F$ are $E_{F,Cu} = 7$ eV, $m_{e,Cu}^* = 1.3 m_e$, $E_{F,Ta} = 5.54$ eV, and $V_{F,Ta} = 1.79 \times 10^6$ m/sec,[24,25] leading to $g_{\uparrow\downarrow,Cu} / g_{\uparrow\downarrow,Ta} = k_{F,Cu}^2 / k_{F,Ta}^2 \approx 1.88$. There is a small difference in results obtained from this analysis (1.88) as compared to that from ISHE measurements (~2.2) or FMR measurements (~2.1). It must be noted that in this simple analysis, the spin scattering at Ta/Cu interface is neglected. However, if the spin current dephasing at the Ta/Cu interface is taken into account in our analysis, the calculated enhancement factor will be smaller than 1.88, and the discrepancy between this model and experimental results will further increase.

To complement the Cu interlayer thickness dependent studies, we have also conducted spin pumping induced ISHE measurements on devices with different Ta thicknesses, which allow us to estimate the spin diffusion length $\lambda_{Ta}$ and spin Hall angle $\theta_{sh}$ in Ta. Using Eq. (2), $J_c = \theta_{sh}(2e/\hbar)J_s$, and $V_{ISHE} = J_c(wd_{Ta})R$, one can obtain

$$\frac{V_{ISHE}}{R} = \zeta\left(\theta_{sh}wd_{Ta}\right)\left(\frac{2e}{\hbar}\right)\frac{\hbar g_{r\uparrow\downarrow}\gamma^2 h_{rf}^2\left(M\gamma + \sqrt{M^2\gamma^2 + 4\omega^2}\right)}{8\pi\alpha^2\left(M^2\gamma^2 + 4\omega^2\right)}\frac{\lambda_{Ta}}{d_{Ta}}\tanh\left(\frac{d_{Ta}}{2\lambda_{Ta}}\right) \quad (3)$$

where $V_{ISHE}$ is the spin pumping induced ISHE signal and $R$ is the resistance measured from Ta/Py devices. $h_{rf}$ is obtained from Ampere's law using $2L_{sg}h_{rf} = \sqrt{2P_{inp}/R}$,[26] where $P_{inp}$ is the input power of 15 dBm (31.62 mW), $L_{sg}$ is 60 $\mu$m, and $R$ is 50 $\Omega$. Thus $h_{rf}$ is calculated to be 3.71 Oe. The factor $\zeta$ in Eq. (3) takes into account the fact that only a part of the sample contributes to spin pumping, *i.e.* the area just below the ACPS strip and the area traversed by propagating spin waves. $\zeta$ can be obtained by noting that ACPS signal line is 60 $\mu$m and ground line is 180 $\mu$m. The $h_{rf}$ below the ground line is 3 times smaller and hence the contribution from ground line should be divided by 3 times. Thus the effective length of device that contributes to spin pumping is $60 + 180/3 = 120$ $\mu$m. The length of device is 800 $\mu$m, thus $\zeta$ can be calculated to be $120/800 = 0.15$. If the contribution of spin waves is also taken into account, the effective length of device will be $60 + 2\ell + (180 + 2\ell)/3$, where $\ell$ is spin wave propagation length of about 25 $\mu$m[27]. Thus the value of $\zeta$ is 0.233, when spin wave contribution is also taken into account. In Fig. 3(a), we have plotted the resistance of Ta/Py devices with different Ta thicknesses. The Ta resistivity is determined to be 130–160 $\mu\Omega\cdot$cm. In Fig 3(b) the ratio of $V_{ISHE}$ and $R$ is plotted as open squares, and is fitted with Eq. (3), resulting in $\lambda_{Ta}$ = 8 nm, and $\theta_{sh}$ = 0.021 or 0.033 for analysis with or without spin wave contribution, respectively. The values of $\theta_{sh}$ are much smaller than the value of 0.15 obtained by the spin torque ferromagnetic resonance method[28]. Similar disagreements have also been found for the case of Pt, in which various values of spin Hall angles in the range 0.0037 – 0.08 have been reported[8,29,30].

Our experimental observation of enhanced $\alpha$ due to the presence of Cu interlayer in Ta/Cu/Py is in contrast with the previous report in which a Cu interlayer was found to decrease the FMR linewidth in Pt/Cu/Py devices[31]. The difference may come from the choice of materials in devices used in the experiment. In order to confirm the effect of Cu interlayer in Pt-based devices, we have conducted spin pumping induced ISHE measurements as well as FMR measurements in Pt/Cu/Py devices. The Pt and Py thicknesses are 10 nm and 20 nm, respectively, while the thickness of Cu interlayer is varied from 0 to 5 nm. Figure 4(a) shows that the ratio of $V_{ISHE}$ and $R$ decreases by about 20% in the presence of Cu interlayer in Pt/Cu/Py devices. The FMR measurements results plotted in Fig. 4(b) also show a decrease in the linewidth in Pt/Cu/Py devices in the presence of Cu interlayer. In Fig. 4(c) the value of $g_{\uparrow\downarrow}$ is plotted as a function of Cu interlayer thickness. It is clear that the effective spin mixing conductance is suppressed in the Pt/Cu/Py system as compared to that of Pt/Py. This result shows that the effect of Cu interlayer on $g_{\uparrow\downarrow}$ is indeed material dependent (Ta versus Pt). Using the $E_F$ of 9.74 eV for Pt and $m^*_{e,Pt} = 13 m_e$,[32,33] the ratio of spin mixing conductance of Cu and Pt can be calculated to be 0.27, which is in qualitative agreement with our results that the spin pumping efficiency decreases in Pt/Cu/Py devices in the presence of a Cu interlayer.

In summary, we have investigated the effect of interface on the efficiency of spin pumping process. Spin pumping induced spin current density obtained from Ta/Cu/Py devices is larger than that of Ta/Py devices. However, the Cu interlayer decreases the spin pumping efficiency in Pt/Cu/Py. The enhancement in the FMR linewidth also confirms an increase of the induced spin current density in Ta/Cu/Py devices, and decrease in Pt/Cu/Py devices. We suggest that the increase in the spin current density in Ta/Cu/Py devices is because of an increase in the spin mixing conductance in the presence of a Cu interlayer. In case of Pt/Cu/Py devices, the spin mixing conductance decreases in the presence of Cu interlayer, and hence the spin current density decreases. These experimental results show that the spin current induced by spin pumping can be tuned by using a metallic interlayer inserted at the NM/FM interface. In addition, we have evaluated the spin diffusion length to be 8 nm in Ta and spin Hall angle to be in the range 0.021-0.033.

This work is partially supported by the Singapore National Research Foundation under CRP Award No. NRF-CRP 4-2008-06 and Singapore Ministry of Education Academic Research Fund Tier 1.

11

**Figure captions**

**Fig. 1** (a) A schematic representation of the measurement setup. A rectangular Ta/Cu/Py layer is patterned. An ACPS line is patterned on top of the SiO$_2$ layer. A signal generator (SG) is connected to the ACPS line and a voltmeter is connected across the multilayer for measuring the spin-pumping signal. (b) Cross section of the device. The layer stack is Ta or Pt (10 nm)/Cu (0-10 nm)/Ni$_{81}$Fe$_{19}$ (20 nm)/SiO$_2$ (30 nm).

**Fig. 2** (a) The measured spin pumping signal (open squares) at 4 GHz as a function of applied bias field $H_b$ for Ta/Py device, and the corresponding fit (solid line). (b) Resonance fields (open squares) as obtained from experiments, and a curve fit by Kittel formula (solid line). (c) The resistance of Ta/Cu/Py devices with different Cu thicknesses. (d) $V_{ISHE}/R$ (~ $J_S$) for Ta/Cu/Py devices with different Cu interlayer thicknesses, normalized with respect to Ta/Py device. (e) The measured FMR signals at $H_b$ = 126 Oe as a function of frequency centered at a resonance frequency ($f_0$), for Ta/Cu/Py devices with different Cu thicknesses. The linewidth of FMR signal is a measure of Gilbert damping $\alpha = \Delta\omega/\gamma$. (f) The effective spin mixing conductance calculated from $g_{r\uparrow\downarrow} = 4\pi M \gamma d_{Py}(\Delta\alpha)/g\mu_0\mu_B$ using an increase of Gilbert damping ($\Delta\alpha$).

**Fig. 3** (a) The resistance of Ta/Py devices for different Ta thicknesses. (b) The measured $V_{ISHE}/R$ (open squares) and a fit (solid lines) for Ta/Py devices with different Ta thicknesses.

**Fig. 4** (a) $V_{ISHE}/R$ for Pt/Cu/Py devices with different Cu thicknesses, normalized with respect to Pt/Py device. (b) The measured FMR signals at $H_b$ = 126 Oe as a function of frequency for Pt/Cu/Py devices with different Cu thicknesses. (c) The effective spin mixing conductance as a function of the Cu interlayer thickness.

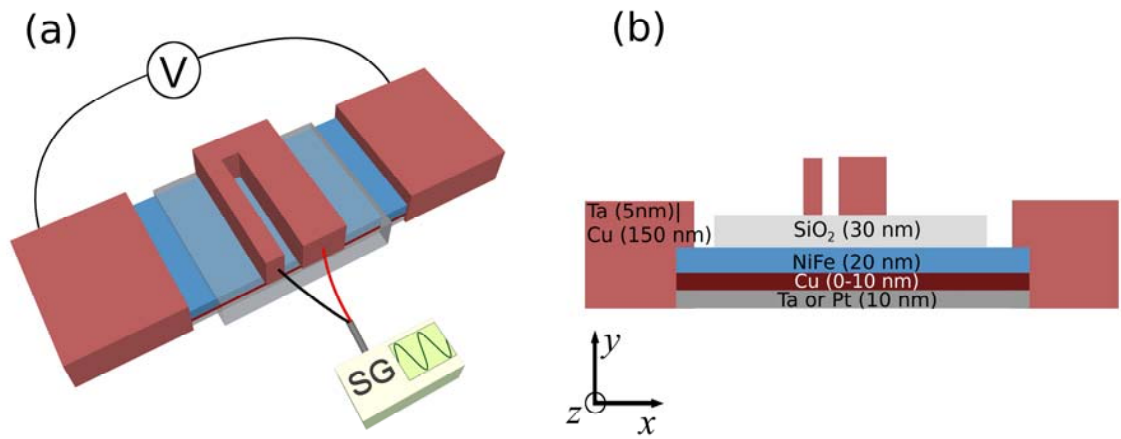

Fig. 1

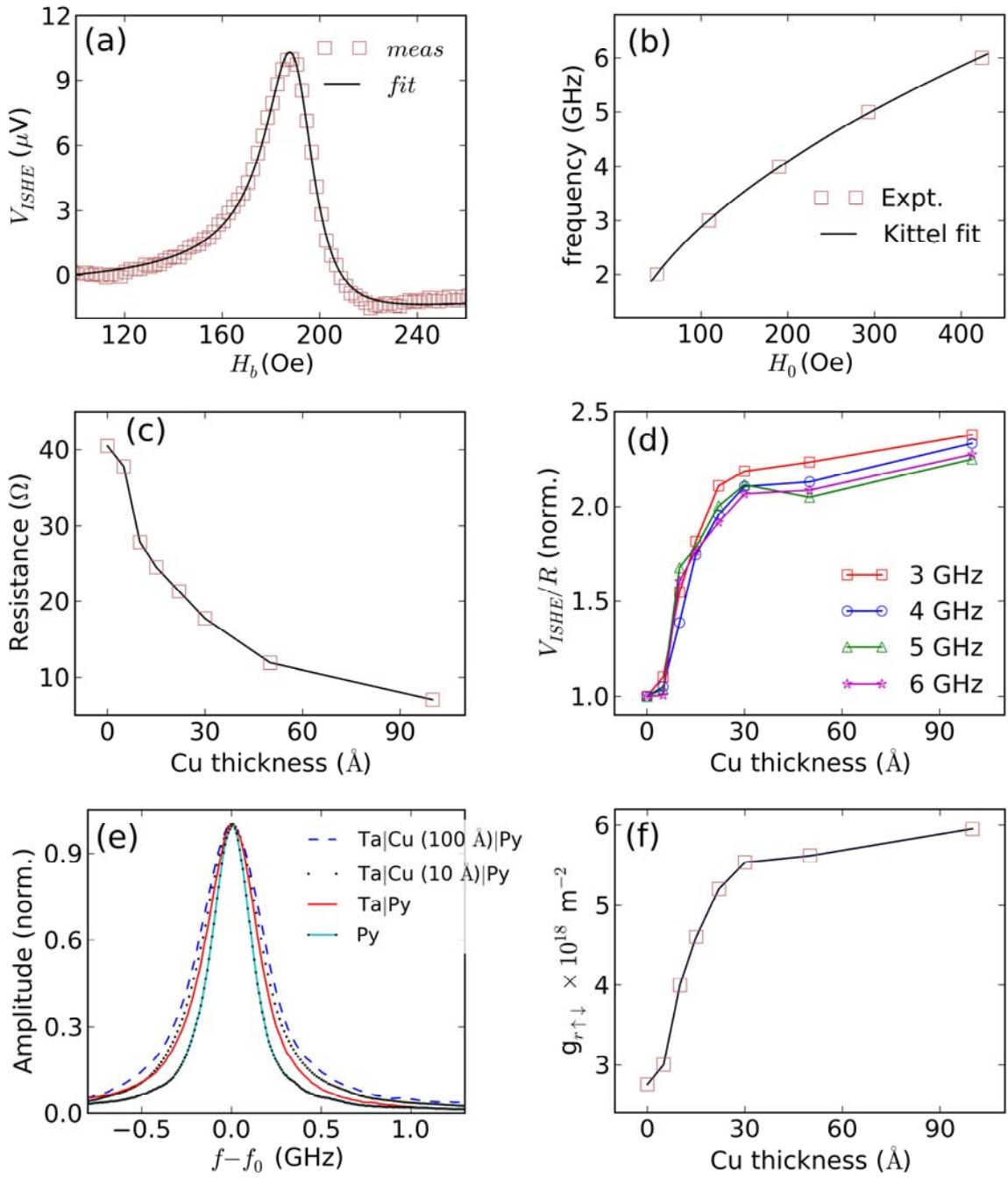

Fig. 2

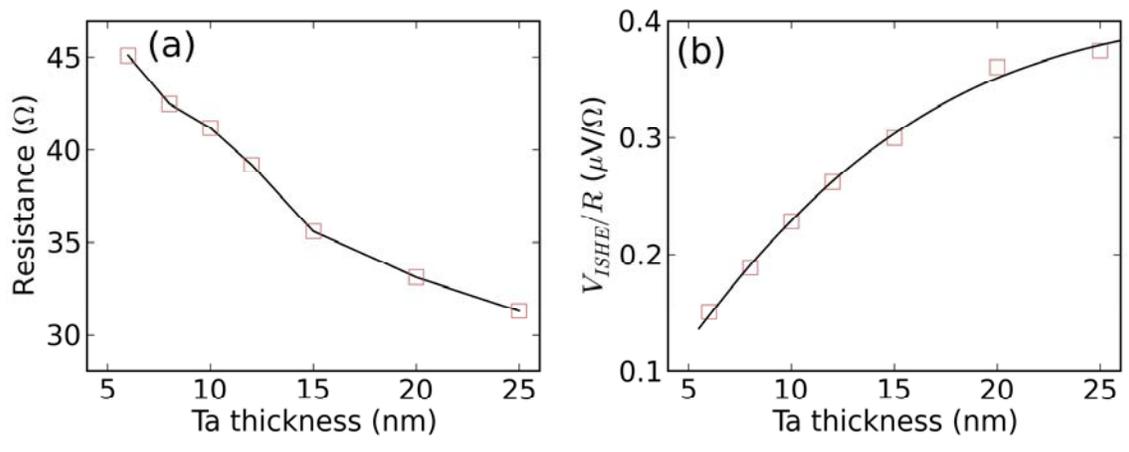

Fig. 3

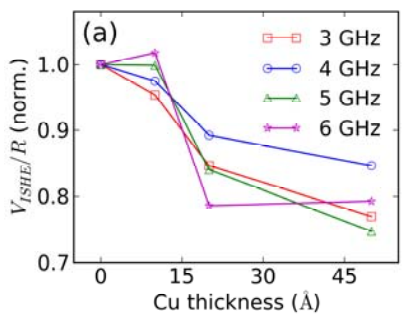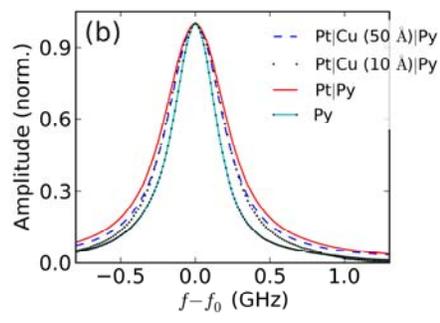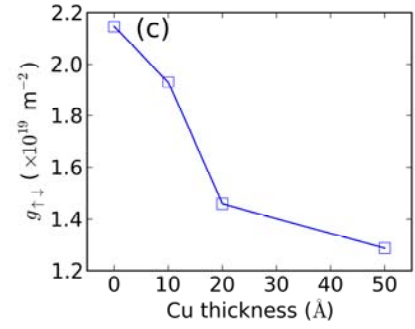

Fig. 4